%% file: main.tex
\documentclass[format=acmtog,screen=true ]{acmart}
\acmSubmissionID{379}

\usepackage{booktabs} 
\usepackage{colortbl}
\usepackage{graphicx}

\usepackage[normalem]{ulem}
\usepackage[inline]{enumitem}
\usepackage{multirow}
\usepackage{multicol}
\usepackage{algorithm}
\usepackage[noend]{algpseudocode}

\algnewcommand\algorithmicforeach{\textbf{for each}}
\algdef{S}[FOR]{ForEach}[1]{\algorithmicforeach\ #1\ \algorithmicdo}

\acmPrice{15.00}

\copyrightyear{2017}
\acmYear{2017}
\acmConference{Conference Name}{Conference Date and Year}{Conference Location}
\acmDOI{10.1145/8888888.7777777}
\acmISBN{978-1-4503-1234-5/17/07}

\setcitestyle{square}

\renewcommand\footnotetextcopyrightpermission[1]{} 

\begin{document}

\title{HMLFC: Hierarchical Motion-Compensated Light Field Compression for Interactive Rendering}

\author{Srihari Pratapa}
\affiliation{%
 \department{Department of Computer Science }
 \institution{Uinversity of North Carolina at Chapel Hill}}
\email{psrihariv@cs.unc.edu}

\author{Dinesh Manocha}
\affiliation{%
\department{ Department of Computer Science and Electrical \& Computer Engineering }
 \institution{University of Maryland at College Park}}
\email{dm@cs.umd.edu}



\begin{abstract}
We present a new  motion-compensated hierarchical compression scheme (HMLFC) for encoding light field images  (LFI) that is suitable for interactive rendering. Our method combines two different approaches, motion compensation schemes and hierarchical compression methods, to exploit redundancies in LFI. The motion compensation schemes capture the redundancies in local regions of the LFI efficiently (\textit{local coherence}) and the hierarchical schemes capture the redundancies present across the entire LFI (\textit{global coherence}). Our hybrid approach combines the two schemes effectively capturing both \textit{local} as well as \textit{global} coherence to improve the overall compression rate. We  compute a tree from LFI using a hierarchical scheme and use phase shifted motion compensation techniques at each level of the hierarchy. Our representation provides random access to the pixel values of the light field, which makes it suitable for interactive rendering applications using a small run-time memory footprint. Our approach is GPU friendly and allows parallel decoding of LF pixel values. We highlight the performance on the two-plane parameterized light fields and obtain a compression ratio of 30--800$\times$ with a PSNR of 40--45 dB. Overall, we observe a $\sim$2--5$\times$ improvement in compression rates using HMLFC over prior light field compression schemes that provide random access capability. In practice, our algorithm can render new  views of resolution $512\times512$ on an NVIDIA GTX-980 at $\sim$200 fps. 
\end{abstract}





\maketitle

\input{intro.tex}
\input{background.tex}
\input{method.tex}

\input{Impl.tex}
\input{results.tex}
\input{concl.tex}
\input{main.bbl}

\end{document}

%% file: intro.tex
\section{Introduction}
\label{sec:intro}

Virtual reality (VR) is being increasingly used for immersive multimedia experiences and telepresence applications. To achieve a high degree of presence in VR, we need to generate high-fidelity renderings of real world scenes at interactive rates. Photo-realistic renderings increase the sense of immersion in real world scenes and provide artistic, life-like experiences in VR\footnote{MIT Technology Review: VR is still a novelty, but Googles light-field technology could make it serious art. \url{https://goo.gl/F79udn}}. The plenoptic function (7D) describes the total flow of light through all the points in space~\cite{plenoptic}. Light Fields (LF) are a low-dimensional (4D or 5D) function of the plenoptic function that capture the radiance of the light rays over a specific region of space. Yu~\cite{LFYu2017} outlines the emergence of light fields and lists the advantages of using LF technology to generate high-quality content  for VR applications. 


Levoy \& Hanrahan~\cite{LFLevoy96} and Gortler et al.~\cite{LFGortler96} describe a 4D parameterized LF and practical approaches for capturing and rendering static scenes using 2D image samples. To generate photo-realistic renderings from different viewpoints, such image-based-rendering (IBR) techniques need large amounts of data to be captured, which is a major issue for interactive applications. The number of image samples required for a good quality rendering using LF is generally in the order of tens of thousands~\cite{LFsampchai2000sampling}. The data sizes of the sampled LF vary from hundreds of MB~\cite{LFLevoy96} to hundreds of GB~\cite{LFSamplin2000number, LFMichelangelo} depending on the scene complexity, sampling rate, and sampling resolution. For 360\textdegree \ panoramic light fields~\cite{WelcomeLF} the LF data-sizes are close to 4--6 GBs. Therefore, compressing the LF is necessary for storing, transmitting, and interactive rendering.


The LF-based rendering algorithms involve retrieving pixel values from LFI and interpolating the pixel values to compute a new view. The pixels required for computing the new view may be located in different regions of different LFI. Therefore, for real-time LF rendering, the relevant portions of uncompressed LFI should be present in the local memory. To render a new view, only a portion of data is required from the entire LFI. As a result, we do not need the entire uncompressed LFI in memory, as it may result in memory bottlenecks. During rendering pixel data is continuously fetched from memory, and in real-time systems, with limited bandwidth (mobile and untethered AR/VR), can cause a significant performance bottleneck~\cite{PVRTC}. Random access compression schemes help in mitigating the memory and bandwidth bottlenecks. Random access compression schemes of LFI have two main properties: ($1$) selective decoding of only the required data; ($2$) allowing fast hardware decompression. To enable interactive LF rendering applications, it is necessary to develop LFI compression schemes that maintain the properties of random access compression schemes as well as provide good compression rates.


Prior LFI compression schemes can be broadly categorized into hierarchical schemes and motion compensated schemes. Hierarchical compression approaches for LFI compute a tree (parent, child dependencies) from the LFI using image transformations and create levels of hierarchy~\cite{LFPeter01,RLFC}. Motion compensation methods capture redundancies in nearby LFI by using a pair of motion vectors  or  disparity values~\cite{zhang2000compression,WelcomeLF}. The hierarchical schemes and motion compensation schemes exhibit different characteristics in terms of  capturing the redundancies across the LFI.

\textbf{Main Results:} We present a new motion compensated hierarchical compression scheme (HMFLC) for encoding LFI for interactive rendering. Ours is a hybrid method that combines two different random access compression approaches to maximize the redundancies captured across the LFI. 
The first class of methods is motion compensation schemes in which the redundancies present in the small regions of the LFI are efficiently captured using extensive search based techniques. The other class of methods is hierarchical compression approaches in which image manipulation and transformation techniques are applied to the entire LFI to capture redundancies across  the LFI in a global manner. We use a hierarchical light field compression approach to capture the redundancies in a global fashion and then apply  phase shifted motion compensation to various levels of the hierarchy. We apply motion compensation to all the levels of the hierarchy by selecting a set of reference frames at each level, creating a new motion-compensated hierarchy. The tree structure computed in the underlying hierarchical scheme is maintained after applying motion compensation at each level. After motion compensation, the amount of data in the levels of the hierarchy is reduced by a significant factor leading to a higher compression rate. We also a present simple and fast scheme to decompress the light fields and use them for interactive rendering on commodity GPUs.
The main contributions of our approach include:

\begin{enumerate}
    \item A novel compression approach combining two different schemes (motion-compensation and hierarchical schemes) for LFI compression to achieve  better compression performance (Section-~\ref{sec:method});
    \item New phase shifted motion-compensation technique suitable for the properties of the images computed in the hierarchy (Section-~\ref{sec:impl});
    \item A hybrid compression scheme (HMLFC) that provides many benefits including random access, progressive decoding, and parallel decompression on commodity hardware (Section-~\ref{sec:impl}).
\end{enumerate}
Our compression algorithm, HMLFC, provides a 2--5$\times$ improvement in compression rate for similar compression quality compared to prior hierarchical schemes as well as motion compensation schemes that provide random access capability (Section-~\ref{sec:results}).  The decompression memory overhead and decompression time overhead due to our hybrid combination is minimal. We can render new views at a resolution of $512\times512$ using an NVIDIA GTX-980 at $\sim$200 fps (Section-~\ref{sec:results}).


%% file: background.tex
\section{Prior Work}
\label{sec:back}

In this section, we give a brief overview of prior work on light field rendering and compression algorithms.

\subsection{Light Field Rendering}

The plenoptic function describes the flow of light in space. Adelson \& Bergen~\cite{plenoptic} use the term plenoptic function (7D) to describe the light intensity $(L)$ at any point $(x, y, z)$ and orientation $(\theta, \phi)$ in free space at any given time $(t)$, and over a range wavelengths $( \lambda)$ in the visible spectrum: $L = P(x, y, z, \theta, \phi, t, \lambda)$. Levoy \& Hanrahan~\cite{LFLevoy96} and Gortler et al.~\cite{LFGortler96} describe a low-dimensional (4-D) form of the plenoptic function, called \textit{light field} as a set of outgoing light rays from a static object or scene. Levoy \& Hanrahan~\cite{LFLevoy96} use two parallel planes described by $(u,v)$ and $(s,t)$, a two-plane parameterization, to describe the 4-D function. Several other low-dimensional parameterizations such as the spherical~\cite{LFSphereihm1997, WelcomeLF} or the unstructured~\cite{LFdavis2012unstructured} have been proposed to describe and capture the plenoptic function. In all the parameterizations, the light rays are captured by densely sampling 2-D camera images from multiple viewpoints around the scene. New views from arbitrary positions in space are generated by interpolating the captured light rays (pixel values). High sampling rates (orders of tens of thousands) are required to achieve photo-realistic reconstructions, which need huge amounts of data to store the captured images~\cite{LFsampchai2000sampling}.

\subsection{Light Field Compression}

A large amount of image data is needed for LF rendering and it creates a bottleneck for interactive applications. LF compression schemes are used to transmit and store LFI for rendering. JPEG Pleno~\cite{JPEGPleno} has been launched by the JPEG standards committee with the goal of establishing standards for the broader adaptability of 4D LF applications. Several compression schemes have been proposed to handle the image data problem in LF rendering. We categorize the existing schemes into two types: \textit{hierarchical compression schemes}, which apply image transformations and image manipulations (wavelet, image warping, and arithmetic manipulations)  to the original LFI and build hierarchical structures that exploit redundancies; \textit{motion compensated compression schemes}, which use standard techniques similar to MPEG video compression (motion vectors) or disparity compensation to capture redundancies. In addition to these two categories, Levoy \& Hanrahan~\cite{LFLevoy96} use vector quantization (VQ) to compress the LFI using a 4D dictionary. The compression rates attained using VQ are around 10:1 to 20:1.  A survey of compression schemes for LFI is presented in Viola et al.~\cite{viola2017comparison}.

\subsubsection{Motion-Compensated Compression Schemes:}

We further categorize motion compensated schemes into two sub-categories: \textit{high efficiency schemes}, which provide very high compression ratios without random access properties and \textit{random access schemes}, which enable interactive rendering by allowing random access to the pixel values of the LFI.

\textit{High-efficiency schemes:} The primary approaches in LFI compression adopt methods similar to image and video compression methods.  These approaches apply techniques such as motion-vector compensation, domain transform (DCT, 
wavelet), and image-warping to exploit redundancies among the LFI. In the case of two-plane 4D parameterized LF, the light rays are sampled using uniform camera motion between adjacent samples. Using this observation,
Girod
et al.~\cite{LFGirod03}; Jagmohan et al.~\cite{LFjagmohan2003compression}; and Magnor \& Girod~\cite{LFmagnor2000} describe methods that use a single disparity value instead of a pair of motion-vector values to encode the LFI predictively. The compression ratios achieved are close to 100:1 to 200:1.
Moreover, these methods do not provide random access capabilities. Very high compression efficiency schemes that provide compression ratios of 100:1 to 1000:1 have been proposed~\cite{LFchen18, LFliu16pseudo, LFperra2016high}. In 
Liu et al.~\cite{LFliu16pseudo}, the grid of LFI is first processed to arrange them in sequential order to 
get an optimal pseudo-temporal ordering that maximizes when compressed using HEVC encoding. Chen et al.~\cite{LFchen18} process the LFI using predictive and image-warping methods from which from a small set of key-views are selected and the rest of the LFI are predicted using the key-views. After the pre-processing, the images are temporally ordered using the method in Liu et al.~\cite{LFliu16pseudo} and then compressed using HEVC. Techniques that use additional information about scene geometry and characteristics in addition to image-warping techniques are presented in Chang et al.~\cite{LFchang2006light}. Image homography is used to warp the LFI onto a fixed set of reference images to find redundancies in Kundu~\cite{LFKundu12}, yielding compression rates of 10:1 to 50:1. Although some of the above methods provide large compression ratios, they fail to address the problems of random access and heavy memory consumption for interactive rendering. These methods are efficient for transmitting and streaming LF data over the internet but require the entire LF to be decoded for rendering. 
\\

\textit{Random Access schemes:} The method in Zhang \& Li~\cite{zhang2000compression} is the first approach that uses motion compensation and provides random access capabilities for LFI compression. They describe a multi-reference, frame-based motion compensation approach that provides compression ratios of 80:1. Overbeck et al.~\cite{WelcomeLF} present a scheme for compressing 360\textdegree \ panoramic light fields captured using their LF capturing system. They achieve 
compression ratios of 40:1 to 200:1 on the complex panoramic LFI datasets.

\subsubsection{Hierarchical Compression Schemes:}
Peter \& Stra{\ss}er~\cite{LFPeter01} present a 4D wavelet hierarchical scheme for compressing LF that provides random access. This method uses 4D Harr wavelets to transform the LFI into wavelet coefficients and organizes the coefficients into a tree structure. They attain compression rates of 20:1 to 40:1 and their method makes assumptions about the scene captured in the light field. Pratapa \& Manocha~\cite{RLFC} present a hierarchical compression scheme that is based on computing representative and residual views at each level of the hierarchy to exploit redundancies across the LFI. The top-level images of the hierarchical tree capture the redundant common details among the LFI and the other levels of the tree store the low-level, high-frequency details of the LFI. Their method obtains compression rates of 20:1 to 200:1, provides random access to the compressed stream, enables progressive decompression, and supports fast hardware decoding. Magnor \& Girod~\cite{LFHierarchyMagnor} describe a hierarchical predictive-based encoding scheme using disparity maps. In Magnor \& Girod~\cite{LFHierarchyMagnor}, an explicit hierarchical tree is not constructed, but a hierarchical relationship among the LFI is established by iteratively dividing the LFI into sub-quadrants. This method provides compression rates of around 400:1, but it does not provide random access capability for interactive rendering.

%% file: method.tex
\section{Motion Compensation \& Hierarchical Compression: Coherence}
\label{sec:method}

 \begin{figure*}[t!]
 \centering
 \includegraphics[width=1.0\textwidth, keepaspectratio=true]{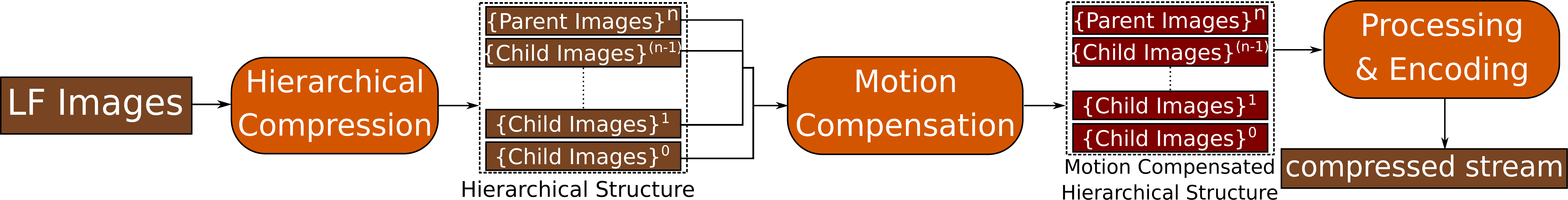}
 \caption{Overview of our HMLFC compression pipeline: The compression pipeline consists of different stages. In the first stage, a hierarchical compression scheme is applied to LFIs to compute levels of new, transformed parent and child images. In the next step, all the levels of the computed hierarchy are processed using a motion compensation scheme to compute a new motion-compensated hierarchy. In the final stage, the motion compensated hierarchy is further processed and encoded to generate the compressed bit stream.}
 \label{fig:hybrid_pipeline}

 \end{figure*}

In this section, we present high-level descriptions of motion compensation and hierarchical compression methods for LFI that provide random access. We discuss the advantages and limitations of each of these approaches and motivate the design of our hybrid compression scheme. In the following discussion, we assume 4D two-plane parameterization, though our approach can be extended to other LF parameterizations. 

For a set of light field images captured using two-plane parameterization, redundancies are present across all the captured LFI. The amount of coherence between two captured light field images varies based on the distance between the actual image capture points in the space. Adjacent light field images exhibit higher coherence, while far off images exhibit lesser coherence. We refer to the coherencies that are commonly present across the entire LFI as \textit{global coherencies}. We refer to the coherencies present across the adjacent images of the LFI as \textit{local coherencies}. An example of LFI highlighting local and global coherencies is highlighted in the suppl. material (Section 3).

\subsection{Motion Compensation Methods}

We use the following terminology to give an overview of prior motion compensation schemes used for LF compression.

\noindent  \textit{Reference Images}: The set of images selected from the original LFI that are encoded independently, using standard compression schemes and used as the reference set for encoding the rest of the images in the LFI in the motion compensation schemes.

\vspace*{0.1in}

\noindent \textit{Predictive Images}: The set of images from the original LFI that are associated with a reference image and are encoded using motion compensation techniques. 

\vspace*{0.1in}

 At a high level, motion compensation schemes start by selecting a subset of frames from the LFI as \textit{reference images}. The process of selecting the \textit{reference images} varies depending on the exact compression scheme, as discussed in Section ~\ref{sec:back}. Once the \textit{reference images} are selected, the rest of the LFI are marked as \textit{predictive images}. Each of the \textit{predictive images} is associated with a \textit{reference image} and encoded using motion compensation. The predictive images are divided into non-overlapping rectangular blocks, and each block of pixels is predicted (computed as difference) from the reference image using a pair of motion vectors. Motion compensation schemes typically use an exhaustive search over a large region in the reference image to minimize the residual difference for each block in the predictive images. Therefore, they compute the redundancies between a given reference image and the associated predictive images very efficiently using an exhaustive search.

 The LFI exhibit a large amount of coherence across all the LFI captured. Motion compensation schemes capture \textit{local coherencies} effectively using exhaustive search. Although the motion compensation schemes exploit \textit{local coherencies} (reference and predictive image sets) of the LFI efficiently using an exhaustive search, they fail to capture the redundancies present across the entire LFI in a global fashion (e.g., coherencies across grids in Fig~\ref{fig:hybrid_overview}).

 \subsection{Hierarchical Methods}
 
 We use the following terms to present an overview of the hierarchical approach:

\noindent \textit{Parent Images \& Child Images}: The new sets of transformed images computed using image manipulations and transformations from the original LFI in hierarchical compression schemes. The new sets of transformed images are  parent image sets and children image sets forming hierarchical relationships between the sets, creating a hierarchy.  
 
\vspace*{0.1in}
 
Hierarchical schemes use image manipulation and transformation techniques on the LFI to compute a new set of images capturing the redundancies across the entire LFI. The new set of transformed images is partitioned into two subsets, parent images and child images, creating a hierarchy. The image manipulation and transformations (wavelet transforms, image warping, image filtering, and arithmetic manipulations) used to compute the new set of images and the exact parent-child relationships depend on the particular compression scheme. Typically, the parent images capture the common redundant details across the LFI and the children contain image specific low-level details of the LFI. The parent subset is further processed recursively to compute the next level of the hierarchy.

The primary advantage of  hierarchical methods is that they capture the \textit{global coherencies} across distant images of the LFI, which the motion compensation schemes fail to capture. Due to the lack of an exhaustive search for redundancies, the global redundancies that are encapsulated in the parent images are limited by the image transformation and manipulation techniques employed in the compression scheme. Figure~\ref{fig:hybrid_overview} (b) shows a high-level overview of a hierarchical LFI compression scheme.

\subsection{Challenges in the Hybrid Approach}
Our goal is to develop a hybrid scheme that captures the benefits of motion compression schemes in terms of local coherency and hierarchical schemes in terms of global coherency. We design a hybrid approach that is based on applying an additional layer of motion compensation to a hierarchical representation. To  obtain good compression rates and provide random access capabilities, we need to address these issues:
\begin{enumerate}
    \item  Once the redundancies across the LFI are captured globally, the properties of the parent and children images computed using a hierarchical scheme differ significantly from the properties of typical images or original LFI. To effectively capture the local coherency, we need new motion compensation schemes to account for the properties of the transformed images in the hierarchy for achieving further compression. 
    \item The resulting motion compensation scheme should conserve all the properties and benefits of the underlying hierarchical scheme, including the hierarchy structure and random access capability.
    \item The overhead of the additional costs of decompression after an additional layer of motion compensation should be minimal.
\end{enumerate}

%% file: Impl.tex
\section{Our Method: HMLFC}
\label{sec:impl}
In this section, we describe our novel hybrid compression algorithm that captures local and global coherency and addresses the challenges highlighted above.

\subsection{Overview}
To tackle the limitations of the \textit{motion compensation methods} and \textit{hierarchical methods}, we combine both approaches to capture redundancies in both \textit{global} and \textit{local} fashion, resulting in better compression rates. In other words, we first apply a hierarchical approach to the LFI gathering all the \textit{global coherencies}, then apply an additional layer of compression to the images at each level to capture remaining redundancies using motion compensated search.

\subsection{Exploiting Local and Global Coherence}

 Our approach (HMLFC) uses a hierarchical motion compensation scheme to capture the redundancies present across the entire set of LFI in a global fashion (\textit{global coherence}). Next, we treat the parent images and each of the children subsets at all levels computed from a hierarchical scheme as  separate subsets of images. Our goal is to apply  motion compensation methods to each of the subsets independently and design a new scheme that exploits the properties of these subsets. The application of motion compensation on the children and parent subsets further exploits the local redundancies (\textit{local coherence}) efficiently by using the exhaustive search that the hierarchical methods fail to exploit. 
 

The decoding properties (random access, progressive decoding, and hardware decoding) of our hybrid approach depend mainly on the underlying hierarchical scheme used for computing the hierarchy and the motion compensation scheme. In the following section we present the overview of the underlying hierarchical scheme and the details of the compression and decompression of our hybrid approach.

 

 
 \subsection{RLFC: Hierarchical Compression Scheme}
 
We choose the RLFC hierarchical compression algorithm described in  Pratapa \& Manocha~\cite{RLFC} because it provides random access to the compressed data and allows hardware decompression. In addition to RLFC, we use a novel motion compensation scheme on the levels of the hierarchy. We maintain the random access property of RLFC after this motion compensation step.


 
In RLFC, the LFI are clustered based on the spatial locations of the samples. For each of the clusters, a new image referred to as the \textit{representative key view (RKV)} is computed by filtering all the image samples in the clusters. The RKV images encapsulate common details among all the images in a given cluster, and the set of RKVs from all the clusters forms a new level (parent images) in the hierarchy. After computing the new level of RKVs, the differences between the RKV and the images in the corresponding cluster are computed. The difference images are referred to as \textit{sparse residual views (SRV)} and the new set of SRVs are the child images in the hierarchy. The SRVs are high-frequency images that contain the specific low-level details of the images that are not captured in the RKVs. This process is recursively implemented on the new RKVs until the tree height reaches a user-set level. We refer the readers to  Pratapa \& Manocha~\cite{RLFC} for exact details of the hierarchy and tree structure computed in RLFC.


 \textbf{Notation:} We use the following notation for explaining the approaches: $R_i$ denotes the $i^{th}$ reference image in the motion compensation methods; $P_i^k$ denotes the $k^{th}$ predictive image associated with $R_i$ in the motion compensated methods; $(x, y)$ denotes the motion vector pair in the motion compensation schemes; $B_{P_i^k}$ denotes a block of pixels in the $k^{th}$ predictive image; $B_{R_i}^{xy}$ denotes a block of pixels for motion vectors $(x, y)$ in the reference image $R_i$; $\Delta$ represents the prediction residual error computed between the reference block and the predictive block.

  \begin{figure}[t!]
 \centering
 \includegraphics[width=\columnwidth, keepaspectratio=true]{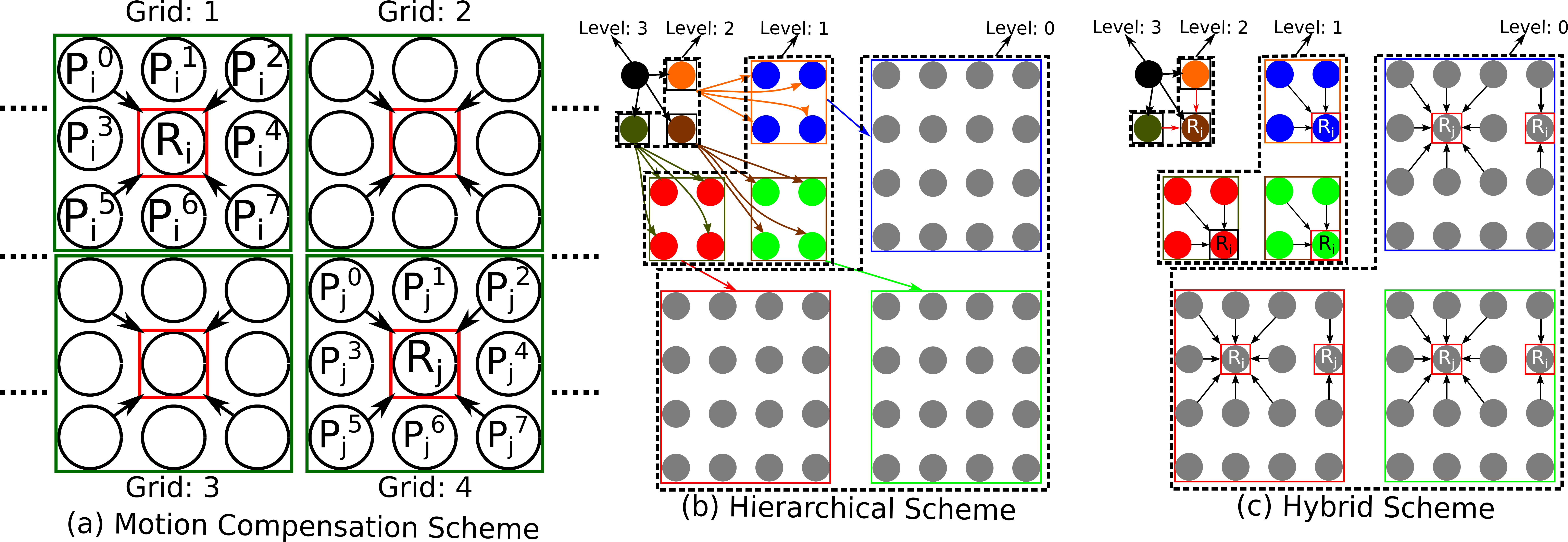}
\caption{An example overview of our hybrid compression scheme. (a) Motion Compensation: The local coherencies are effectively captured in small regions (grids marked in green) of the LFI using an exhaustive search. A set of images from LFI is selected as the reference images, highlighted in red ($R_i$, $R_j$). The rest of the images are marked as predictive images and are compressed from the reference images using motion compensation. (b) The coherencies present across the entire LF are captured using image manipulation and transformations applied to the LFI in a global fashion. The levels of the hierarchy are marked and arrows indicate parent-child relationships in the hierarchy. (c) Our Hybrid Approach: Once the global coherencies are encapsulated using the hierarchical scheme, the additional redundancies at each level of the hierarchy are captured as local coherencies by applying motion compensation at every level. At each level of the hierarchy, reference images ($R_i$, $R_j$) are indicated using a bounding box. The predictive images associated with the reference images are indicated with arrows pointing towards the reference images.}
 \label{fig:hybrid_overview}
 \vspace*{-2.5em}
 \end{figure}

\subsection{ Phase-shifted Motion Compensation}

\begin{figure}[t!]
    \centering
  {\includegraphics[width=\columnwidth, keepaspectratio=true]{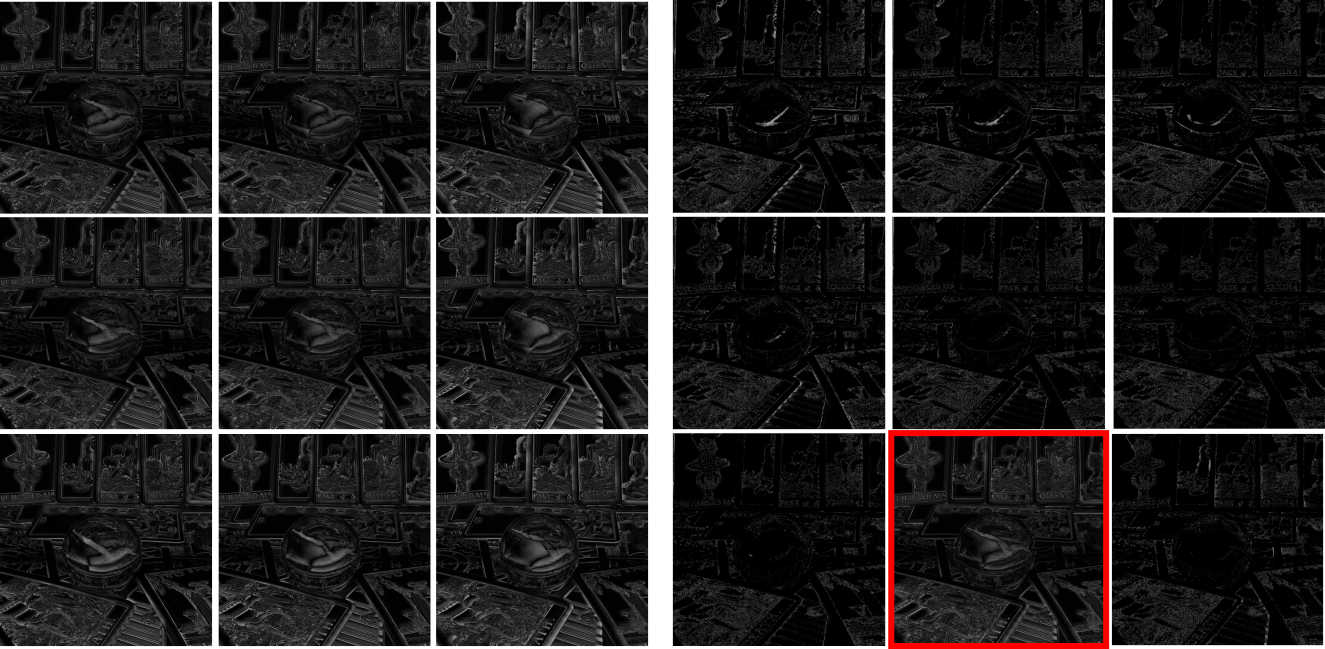}}
    \caption{(left) An SRV image cluster from the RLFC hierarchy from level: 0. It is evident that SRV images are visually similar to each other and exhibit a lot of coherency between them. We exploit the redundancy by applying motion compensation to achieve further compression. (right) The SRV image cluster is shown at left after applying  motion compensation. The reference image is highlighted in red, and the rest of the images are residual difference SRV images after motion compensation. Compared to the original SRV images, the residual difference SRV images are sparser leading to  significantly better compression rates.}
    \label{fig:mv_srv}
\end{figure}

As shown in Fig.~\ref{fig:mv_srv} (left), the SRV images exhibit significant local coherency at each level of the hierarchy. The SRV images are computed as the difference between RKV images at a given level and images in the level below. Due to the difference computation,  the pixels in the SRVs have both negative and positive intensity values. The negative and positive pixel intensity values correspond to the inversions of pixel intensity values across the SRV image signals at a given level. We refer to these inversions as {\em phase-shifts} in the SRV image signals. We present a new phase-shifted motion compensation to capture \textit{local coherencies} in the levels of the hierarchy. These phase shifts in the SRV image signals need to be accounted  while applying motion compensation to the SRV images. More details about the phase shifts that occur in the SRV images are presented in the suppl. material, Sec-1.

For a selected SRV reference image $R_i$ at any given level, let $\lbrace P_i^0, P_i^1, .., P_i^n\rbrace$ denote the set of predictive SRV images associated with $R_i$. Each block ($B_{P_i^k}$) in the predictive $P_i^k$ is motion compensated by searching over a large search window $W$ in the reference SRV image $R_i$ using a pair of motion vectors $(x, y)$. We include the phase shifts in our motion prediction scheme by computing two residual errors for each block (number of pixels in a block: $N$) as follows:


\begin{align}
\vspace*{-2em}
    \Delta_-^{xy} = \sum_{l=1}^{N} \ \lvert B_{P_i^k}(l) - B_{R_i}^{xy}(l) \lvert, \\ 
     \Delta_+^{xy} = \sum_{l=1}^{N} \ \lvert B_{P_i^k}(l) + B_{R_i}^{xy}(l) \lvert. 
\end{align}
For a given pair of motion vectors $(x, y)$,  we compute a subtractive prediction residual error $\Delta_-^{xy}$ and an additive prediction residual error $\Delta_+^{xy}$ to include possible phase shifts between the reference image and the predictive images in a given region.
 \begin{align*}
    \Delta_- = \min_{(x^-, y^-)}( \Delta_-^{xy}) \ \ \forall x, y \in [-W, W], \\ 
     \Delta_+ = \min_{(x^+, y^+)}( \Delta_+^{xy}) \ \ \forall x, y \in [-W, W]. 
\end{align*}
The minimum subtractive prediction residual error $\Delta_-$  and the minimum additive prediction residual error $\Delta_+$ are computed for each block ($B_{P_i^k}$). $(x^-, y^-)$ and $(x^+, y^+)$ are the motion vectors corresponding to the minimum prediction residuals. The final motion prediction residual error $\Delta$ and the corresponding motion vectors are computed as follows:
  \begin{align*}
    \Delta = \min( \Delta_-, \Delta_+), \\
    (x, y) =  \begin{cases}
     (x^-, y^-) & \text{if } \ \ \Delta = \Delta_-, \\
    (x^+, y^+) & \text{if } \ \ \Delta = \Delta_+.
  \end{cases}
\end{align*}
 Next, we perform a replacement step in which the original pixel values in the block $B_{P_i^k}$ in the predictive SRV image ($P_i^k)$ are replaced with predictive residuals of the block. The replacement step modifies the SRV images in the original RLFC tree and computes a new \textit{HMLFC tree}, but the tree structure and the hierarchy remain exactly the same as the original RLFC tree. Figure~\ref{fig:mv_srv} shows the predictive residual SRV images after applying our phase inclusive motion compensation. The predictive residual SRV images computed after motion compensation are much sparser than the original SRV images. Therefore, the predictive residual SRV images can be compressed more significantly without quality loss, resulting in better compression rates. A zoomed-in (16X) visual comparison between an original SRV image and the corresponding predictive residual SRV image after motion compensation is shown in suppl. material, Sec-5. 
 
\subsection{Compressing HMLFC tree}

 YCoCg~\cite{InvertibleYCoCg} color space is used in our implementation to decorrelate the RGB color channels and the chroma channels are sub-sampled. The dynamic range (number of bits to store pixels) of the pixel values is adjusted accordingly to avoid any loss of information due to transformations. The hierarchy computation and motion compensation are performed separately on all the channels. After the motion compensation, the top-level RKV images of the hierarchy are similar to the standard images (RGB) and are compressed using JPEG2000 in the lossless mode.

The compression rate and compression quality of the scheme are controlled by encoding parameters set as user-input to the compression method. The main encoding parameters in the RLFC scheme are \textit{tree height}, \textit{block size}, and \textit{block thresholds}. In addition to the three encoding parameters another important encoding parameter is \textit{search window size} used in the phase-shifted motion compensation  to perform the exhaustive search. The SRV images at all levels of the hierarchy are divided into \textit{block size} non-overlapping rectangular blocks and motion compensation is applied to all the blocks as described in the previous section.

After applying the motion compensation to the hierarchy and computing the HMLFC tree, SRV images (both residual difference SRVs and reference SRVs) are thresholded to discard insignificant data based on the two \textit{block thresholds} set as encoding parameters. For each block in the SRV images, an energy value is computed by summing up the absolute values of the pixels in the block. If the energy is less than the user set threshold the block is marked as insignificant and not stored in the final compression. Due to motion compensation, any losses introduced in the reference SRVs due to thresholding gets propagated to the associated predictive SRVs. To avoid that we use two independent \textit{block thresholds} for thresholding the  motion compensated residual SRVs and reference SRVs.

\subsection{Bounded Integer Sequence Encoding}

In the construction of the hierarchy, we need to perform lossless integer computations, and the final pixel values in all the images of the hierarchy are integer values. BISE~\cite{nystad2012adaptive} presents an efficient way of encoding a sequence of integer values within a fixed range [$0$, $N-1$] and allows for fast random access decoding in constant time with minimal hardware. The straightforward solution for allowing fast hardware random access to the sequence of integers as bit strings of their corresponding binary representations. However, this solution is only optimal when $N$ is a power of two because it uses $\log_2{N}$ bits to store the integer values equivalent to the information present in each integer value.  Besides the simple case when $N$ is a power of two BISE provides an efficient encoding that is close to the information theoretic bounds for other ranges of $N$. The significant blocks in the SRV images after thresholding in the HMFLC tree are encoded using BISE. The resulting formulation is easily supported by the hardware and provides lossless computations.

\subsection{Compressed Stream Structure}

We further process and compress the HMLFC tree and the additional motion vector values computed from the motion compensation step. The HMLFC tree is linearized using breadth-first search (BFS) traversal indexing all the SRVs in the traversal order starting from the top of the tree. The BISE compressed blocks of the SRV images in the tree are arranged in the same BFS linearized order and appended to the compressed stream.  To maintain fast random access property, we extend the application of BISE to encode the motion vector values. Compressing motion vectors using BISE also preserves the fast hardware decompressible property of our stream.

\subsection{Decompression}

\subsubsection{Decoding Procedure:} Decoding a block of pixel values from a particular location from the LFI consists of two main steps: \begin{enumerate*}
    \item Decoding the blocks from the HMLFC hierarchy using tree traversal;
    \item Applying motion re-compensation for the motion compensated blocks.
\end{enumerate*}
At first, the top-level RKV images are decoded and stored in memory. To decode a block of pixels, we use tree traversal procedure and collect the required BISE compressed SRV blocks from the top level to the bottom level. Using the block indices, we infer whether a block belongs to the predictive SRV images in the hierarchy. The original SRV blocks are computed from the predictive residuals and reference SRV images using  motion re-compensation. The motion vector values for the corresponding block are decoded from the random accessible BISE compressed motion vector stream. The final pixel values are computed by combining the SRV pixel values with the corresponding top-level RKV pixel values.
 
\subsubsection{Decompression Memory Overhead:} 
The pixels from the reference SRV images are necessary during decoding to perform the additional motion re-compensation step. To avoid the additional time required for decoding the reference SRV pixels, at the start of the decompression, the reference SRV images at all the levels are decoded and loaded into the memory. The SRV images in the hierarchy are highly sparse in terms of the pixel distribution present in the images (Fig.~\ref{fig:mv_srv}). We use a sparse matrix representation~\cite{SparseMatrix} to store the decompressed reference images in the memory while rendering. The sparse matrix representation reduces the additional run-time memory overhead required for the motion re-compensation step to decode a given block of pixels. Although there is a minor time overhead in reading the pixels from sparse matrices, the overhead is much smaller than the time required for decoding the reference SRV pixels for motion re-compensation. The size of the additional memory overhead depends on two factors:
One of them is a user-set encoding parameter such as the number of levels in the hierarchy and number of reference images in each level. The second factor is the sparsity of the SRV images which depends on details of the scene captured in the light field images.

\subsubsection{Decompression Time Overhead:} The additional layer of motion-compensated step on top of the hierarchy requires decompression and results in additional decompression overhead. This includes tree traversal decoding operations needed to retrieve a block of pixels. Furthermore,  the HMLFC algorithm performs three additional basic operations:
\begin{enumerate*}
    \item Bit manipulations required to decode the corresponding motion vectors;
    \item Loading the bytes of data (pixels) from the reference image in memory into the registers;
    \item Performing arithmetic operations to compute the motion re-compensated block. 
\end{enumerate*} 
In terms of these additional operations required to decode a block of pixels, only the memory load operations are slightly more expensive. In our parallel GPU decoding implementation and experiments (Sec.~\ref{sec:results}), we noticed this overhead to be minimal.

\subsubsection{Random Access for Interactive Rendering:} Random access to the pixel values in the HMLFC tree is guaranteed by the tree traversal decoding operations described (Section 4.8.1). To decode a required pixel value, a block of pixel values corresponding to the required pixel value is decoded. Following that, only the motion vectors corresponding to the predictive residual blocks are retrieved from the BISE compressed motion vector stream. Only a part of the compressed stream is decoded to retrieve required blocks of pixels and the corresponding motion vector values, while the rest of the compressed stream remains intact. Our method also supports parallel decompression of different pixel values from the compressed stream enabling fast GPU decoding. To retrieve a single pixel value of LFI using our decoding, a block of pixel values are decompressed. As a block of pixels are decoded, our method benefits any LF rendering scheme by providing fast access to neighboring pixels for interpolation to compute new views. A set of new views computed for different camera positions and for given LF geometry are shown in suppl. material Sec-8\footnote{Supplementary material link: \url{https://bit.ly/2K2b1Ba}}.

  {\small
\begin{table}[t!]
\centering
\resizebox{\columnwidth}{!}{%
\begin{tabular}{|c|c|c|}
\hline
LF Dataset (Resolution): Size (MB) & \begin{tabular}[c]{@{}c@{}}Compression  \\ rate (bpp)\end{tabular} & PSNR (dB) \\ \hline

Amethyst $(16 \times 16 \times 768 \times 1024): 576$                                               & 0.045                                                              & 40.7     \\ \hline
Bracelet $(16 \times 16 \times 1024 \times 640): 480$                                               & 0.143                                                              & 40.1     \\ \hline
Bunny  $(16 \times 16 \times 1024 \times 1024): 768$                                                & 0.027                                                              & 41    \\ \hline
Jelly Beans $(16 \times 16 \times 1024 \times 512): 384$                                            & 0.029                                                              & 40.5     \\ \hline
Lego Knights $(16 \times 16 \times 1024 \times 1024): 768$                                          & 0.157                                                              & 41     \\ \hline
Lego Gallantry $(16 \times 16 \times 640 \times 1024): 480$                                         & 0.155                                                              & 40.1    \\ \hline
Tarot Cards $(16 \times 16 \times 1024 \times 1024): 768$                                           & 0.68                                                               & 40.3     \\ \hline
\end{tabular}
}
\caption{The compression computed using our HMLFC algorithm  and the quality for several LF datasets from the Stanford light field archive. All the image samples are 24-bit color RGB images. For a similar PSNR quality, the compression rate varies for each LF depending on the details of the scene recorded in the LF.}
\label{tab:overall}
\vspace*{-2.5em}
\end{table}
}

\subsection{Compression Analysis}

We identify two primary properties of the LFI that affect the final compression rate of our method, and we briefly discuss their relationship with the encoding parameters used in our approach.  \begin{enumerate*}
\item Distance between the captured light field image samples;
\item Details of the scene captured in the light field.
\end{enumerate*}

As the distance between the light field samples increases the disparity for a real-world scene point in the pixel space of the adjacent light images also increases. The RKVs are computed as weighted filtering (pixel-wise) of the close  light field images; as the disparity gets higher, the correlation between the same pixels decreases.
As a consequence, the redundancies captured in the RKVs decrease leading to a decrease in the sparsity of the SRV images and more additional redundancies across the SRV images in a given level of the hierarchy. As a result, for a fixed search window size, as the sampling distance between LFI increases, the resulting bit rate increases. For a scene with extensive details, the sampled light field images contain a lot of high-frequency components. In this case, even for a small capture distance between light field images due to the vast regions of high-frequency components, the resulting SRV images have low sparse regions with large intensity values. For a given block threshold, as the complexity of the scene increases, the resulting compression rate also increases as the number of significant blocks in the SRV images increases. However, the redundancies in the high-frequency components of the SRV images in a level can be captured using a motion compensated search. 

\begin{figure}[t!]
\centering
\includegraphics[width=\columnwidth, keepaspectratio=true]{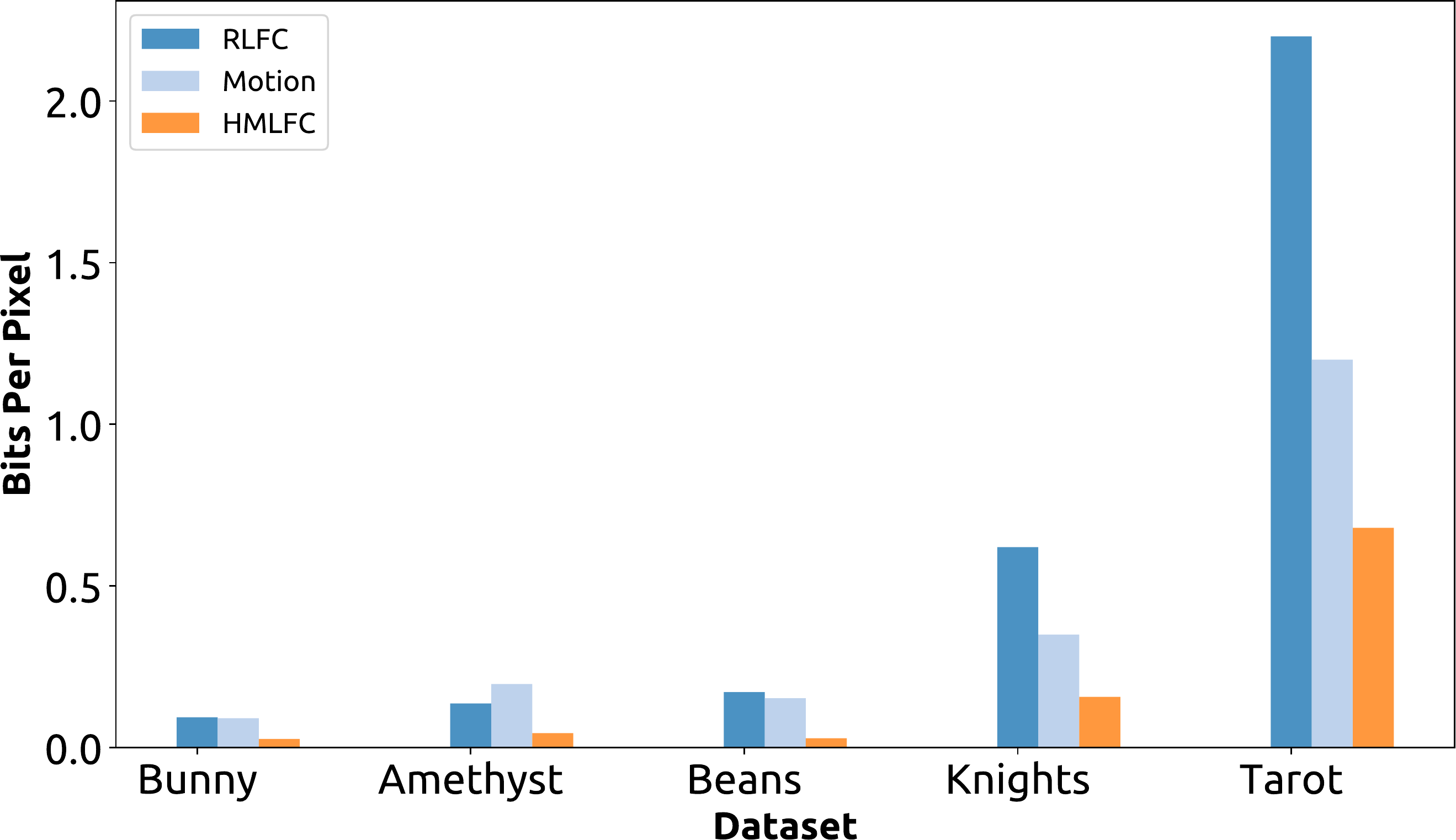}
\caption{We compare HMLFC with RLFC and motion compensation schemes in terms of compression rates (bpp) for several datasets. The datasets are compressed to have a similar compression quality for each of the methods in comparison. Overall, HMLFC improves the compression rate by a factor of $\mathbf{\sim 2-5\times}$, compared to prior schemes.}
 \label{fig:mv_comp}
 \vspace*{-2.5em}
\end{figure}

%% file: results.tex
\section{Evaluation \& Performance Analysis}
\label{sec:results}

We present the results from the evaluation of our hybrid approach and analyze its performance on the Stanford LF archives~\cite{LFLevoy96, LFDatawilburn2005high}. We use peak-signal-to-noise-ratio (PSNR)~\cite{PSNROHM} for quality comparison (suppl. material, Sec-5) and bits per pixel (bpp)  to present the compression rates. We present a comparison of our hybrid method with RLFC~\cite{RLFC} and a motion compensation scheme that enables random access in terms of compression rates and compression quality. The motion compensation scheme is implemented based on Zhang \& Li~\cite{zhang2000compression}.

                           
                               
                           
                           


In Table~\ref{tab:overall}, the compression rates and PSNR values are shown for different datasets from the Stanford LF archive. For a similar PSNR quality,the compression rates vary from $0.029$ bpp to $0.68$ bpp due to the variation of the details captured in the scenes of the datasets. The encoding parameters are varied across the datasets to achieve  a similar compression quality.

Figure~\ref{fig:mv_comp} shows the comparison of compression rates for several datasets for similar compression quality (variation in PSNR quality $0.5-1.5$ dB) for different methods. In some datasets (Bunny, Amethyst, Jelly Beans) RLFC provides similar or better compression than motion compensation. In other datasets (Lego Knights, Tarot Cards) with complex and high-frequency details, we notice that the motion compensation scheme provides better compression rates. We notice that by combining both approaches, HMLFC achieves better compression in both cases. HMLFC improves the compression rate by $\sim 2-5\times$ compared to RLFC for datasets where RLFC provides better compression rates. For other datasets with complex and high-frequency details, HMLFC improves the compression by $\sim 3-5\times$ compared to RLFC and improves the compression by $\sim 2-3\times$ over motion compensation schemes.

\begin{figure}[t!]
\centering
\includegraphics[width=\columnwidth, keepaspectratio=true]{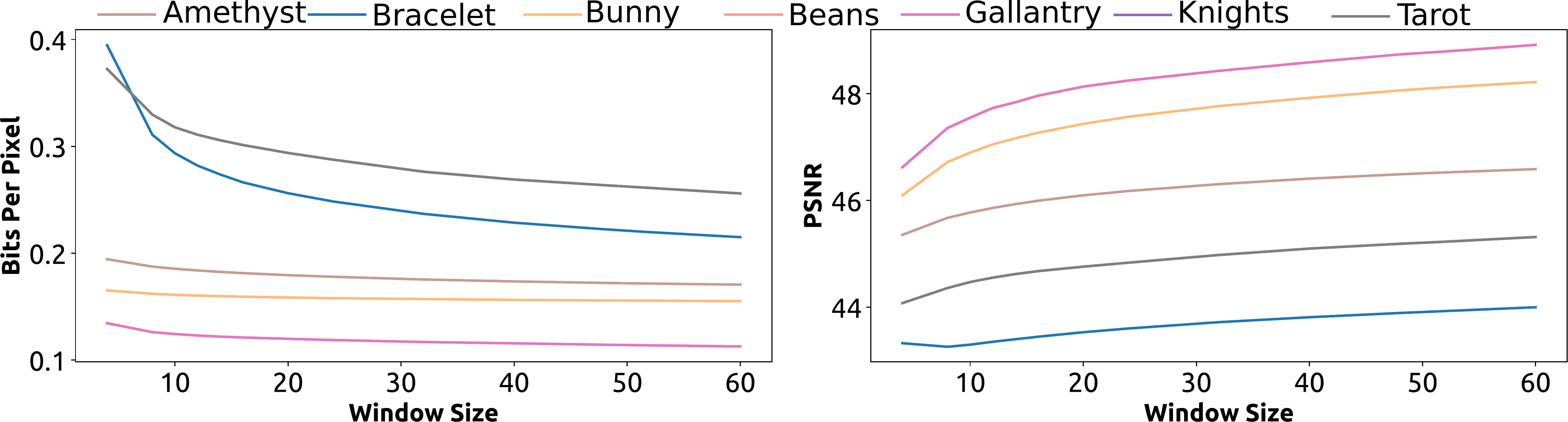}
\caption{We highlight the variation in the compression rates and compression quality of HMLFC with the change in the window size. (left) The increase in the search window size leads to better matching blocks resulting in smaller prediction residual errors and better compression rates. (right) The prediction residual errors are reduced with an increase in the search window size and the resulting compression quality increases. The \textit{block size} is set to 4 and \textit{tree height} is set to 3. The \textit{block threshold} is varied across different datasets to keep the PSNR within a certain range.}
 \label{fig:win_sz_var}
 \vspace*{-3em}
\end{figure}

We analyze the rate-distortion properties of HMLFC by varying the following encoding parameters: \textit{block size}, \textit{block threshold}, and \textit{search window size}. Table~\ref{tab:blksize} shows the variation of the compression rate and resulting quality with a change in the \textit{block size}. Increasing the \textit{block size} with a fixed \textit{block threshold} causes a decrease in the thresholding errors, which results in an increase of PSNR and bpp. Figure~\ref{fig:win_sz_var} shows the effect of varying the window size on the compression rates and compression quality. As the search window size increases, the predictive blocks find better matching blocks in the reference images resulting in a sparser predictive residual. The increase in sparsity of the predictive residuals leads to a reduction in the compression rate. Better matching blocks in the reference images lead to better compression quality and an increase in the PSNR. The coherency between the predictive blocks and reference images is limited to only a certain local region and is diminished beyond a certain search window size. As the window size gets larger than a certain range, we notice that the compression rate and compression quality become saturated. If the spatial distance between the sampled light field images is large, we notice a large benefit in terms of compression (Tarot, Bracelet) as the search window size increases. The results of varying the search window size agree with the compression analysis presented in Section-4.8.

As estimated in the compression analysis in Section - 4.8 and, as presented in Table~\ref{tab:overall} as the details of the contents captured in the scene (example images of the dataset are shown in suppl. material, Sec-2) increase we notice an increase in the bit rate. In the new Stanford LF archive, the Tarot Cards scene is captured with two different sampling distances (small and large) between the light field images. The compression rate on the dataset with the larger sampling distance is $0.68$ bpp for a PSNR of 40.3 dB; on the dataset with smaller sampling distance it is $0.47$ bpp for a PSNR of 40.2 dB.

The variations of compression quality with compression rate for both HMLFC and RLFC are shown in the Figure~\ref{fig:bpp_psnr}. The rate-distortion for  both methods is computed by varying the block threshold. We notice that for different ranges of PSNR, HMLFC achieves better compression compared to RLFC. Visual quality comparison between RLFC and HMLFC is shown in  Figure~\ref{fig:zoomed_in}. The encoding time using our current single-threaded implementation required for compressing varies from 30--90 minutes depending on the input size of the LFI and resolution of the LFI. More compression evaluations of HMLFC (in comparison with RLFC) on datasets Heidelberg LF benchmark~\cite{LFHeidelberg16} in suppl. material, Sec-6. Novel views not present in the original LFI computed for new camera viewpoints are presented in suppl. material, Sec-8. 

\begin{table}[t!]
\begin{tabular}{|c|c|c|c|c|}
\hline
LF Dataset & Metric & \begin{tabular}[c]{@{}c@{}}Block\\ Size: 2\end{tabular} & \begin{tabular}[c]{@{}c@{}}Block\\ Size: 4\end{tabular} & \begin{tabular}[c]{@{}c@{}}Block\\ Size: 8\end{tabular} \\ \hline \hline
\multirow{2}{*}{Amethyst} & PSNR & 38.7 & 43.69  & 48.35\\ \cline{2-5}
                          & bpp & 0.0592 & 0.106 & 0.707\\ \hline \hline
\multirow{2}{*}{Bunny}    & PSNR & 40.52 & 43.35 & 47.52\\ \cline{2-5}
                          & bpp & 0.0173 & 0.0411 & 0.548\\ \hline \hline
\multirow{2}{*}{Bracelet }& PSNR & 37.03  & 44.15 & 48.79 \\ \cline{2-5}
                          & bpp & 0.033 & 0.35 & 1.108 \\ \hline \hline 
\multirow{2}{*}{Knight}   & PSNR & 38.27  & 43.046  & 47.89 \\ \cline{2-5}
                          & bpp &  0.096 &  0.243 & 1.15 \\ \hline    
\end{tabular}
\caption{The effect of varying the block size on the compression rate and quality is highlighted. Increasing the block size for a fixed \textit{block threshold} reduces the total number of thresholding errors, resulting in an increase of bit rate and PSNR. The \textit{block threshold} is set to 75, the \textit{search window size} is set to 16, and the \textit{tree height} is set to 3.}
\label{tab:blksize}
\vspace*{-3em}
\end{table}

\textbf{Decompression Analysis:} We have implemented a GPU LF rendering (more details in suppl. material, Sec-7) using a basic ray-tracing method to test the implementation of our decompression scheme on an NVIDIA GTX-980. We tested the decompression scheme on the Lego Knights dataset compressed using the following encoding parameters: \textit{block size} 4, \textit{tree height} 3, \textit{search window size} 16. Our method takes $3-8$ milliseconds to generate frames at resolution $512\times512$, depending on the number of blocks decoded per frame. The resulting average frame rate for rendering new views is $\sim 200$ fps. The average frame rate to render new views at resolution $1024\times1024$ is close to $\sim 160$ fps. Although HMLFC involves few additional steps in decoding a block of pixels, it achieves similar frame rates as RLFC. We speculate that the decompression of RLFC is bottle-necked on the number of memory operations required to perform the decoding of a block. The inclusion of a few additional memory operations to perform the extra step of motion re-compensation for decoding HMLFC is negligible on the overall rendering performance. The decompression memory overhead to store the sparse matrix representation of the reference images while rendering is $\sim$800 KB in the  Lego Knights dataset. 

@inproceedings{LFHeidelberg16,
  title={A dataset and evaluation methodology for depth estimation on 4d light fields},
  author={Honauer, Katrin and Johannsen, Ole and Kondermann, Daniel and Goldluecke, Bastian},
  booktitle={Asian Conference on Computer Vision},
  pages={19--34},
  year={2016},
  organization={Springer}
}

%% file: concl.tex
\section{Conclusions, Limitations \& Future Work}
\label{sec:concl}

\indent\textbf{Conclusions:} We present a  novel hybrid compression scheme that combines two prior compression methods, \textit{hierarchical schemes} and \textit{motion compensation schemes}, to encode LFI.  Our approach captures the local and global coherencies in the LFI and improves the compression rate by a factor of $\sim 2-5 \times$ without any significant loss in the compression quality. Our scheme provides random access capability and can be used for interactive rendering on current GPUs.   We have highlighted its benefits on standard benchmarks and observe compression rates of $30-800\times$ with a PSNR of $40-45$ dB.

\textbf{Limitations:} Our approach has some limitations. The primary limitation of the hybrid approach is in designing a suitable motion compensation scheme for the transformed images in the hierarchy. Without  proper motion compensation suitable for the underlying hierarchy the benefits, from the hybrid combination might be limited. Another limitation of our method as pointed out in the results (Sec - 5) is that the compression rate is dependent on the distance between the light field images in the light field samples. In the case of light fields captured with a sparse sampling rate, the performance of our compression scheme is reduced. The current GPUD decoder for our compression scheme is not optimized in terms of the memory operations required for decoding.

 \begin{figure}[t!]
\centering
\includegraphics[width=\columnwidth, keepaspectratio=true]{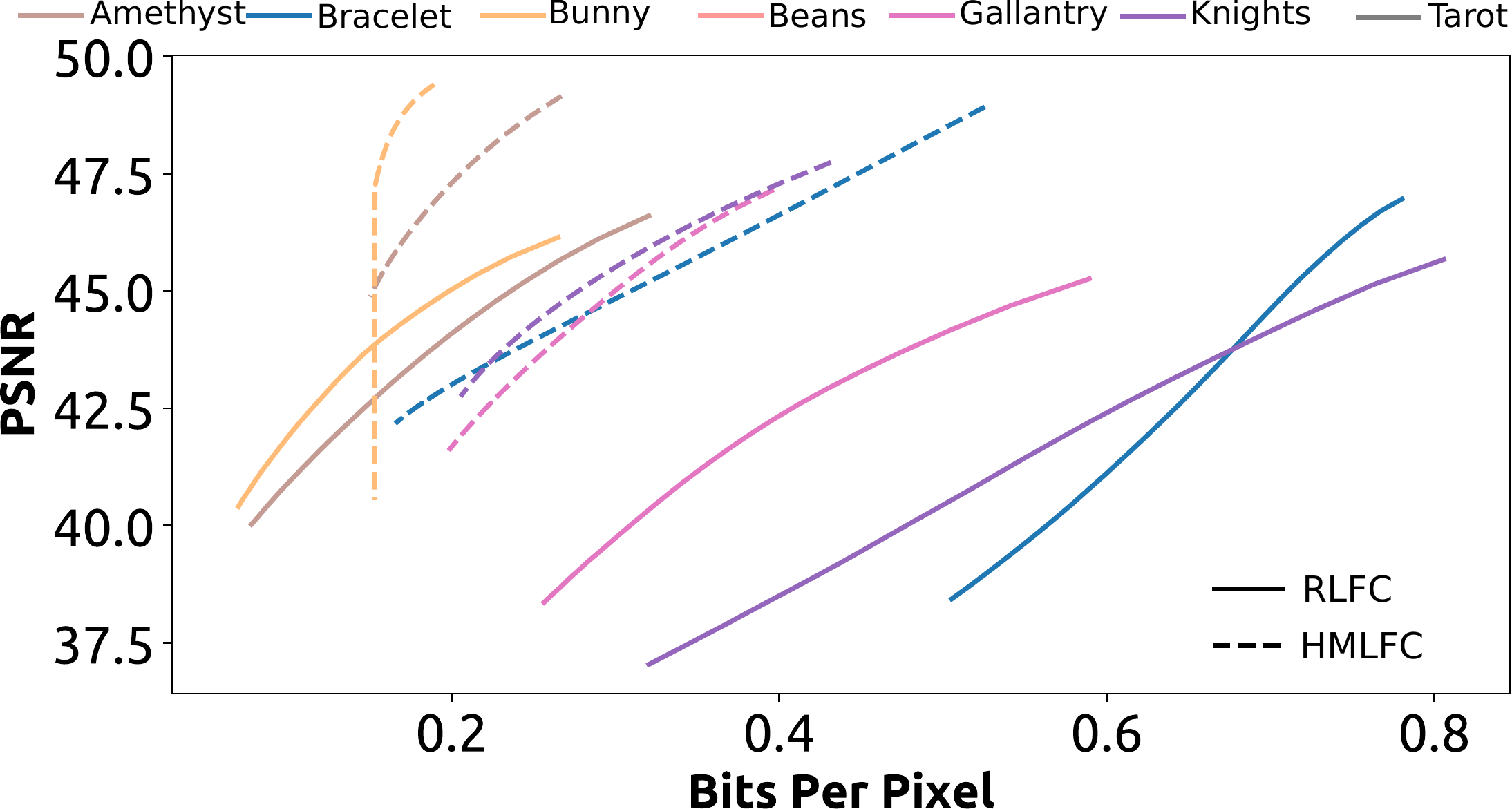}
\caption{The variation of compression quality with bit rate is highlighted for HMLFC and RLFC. HMLFC provides better compression rates for all the datasets over a range of PSNR values. The \textit{block size} is set to 4, \textit{tree height} is set to 3, and the \textit{search window size} is set to 16.}
 \label{fig:bpp_psnr}
 \vspace*{-2.75em}
\end{figure}

  \begin{figure*}[t!]
 \centering
 \includegraphics[width=\textwidth, keepaspectratio=true]{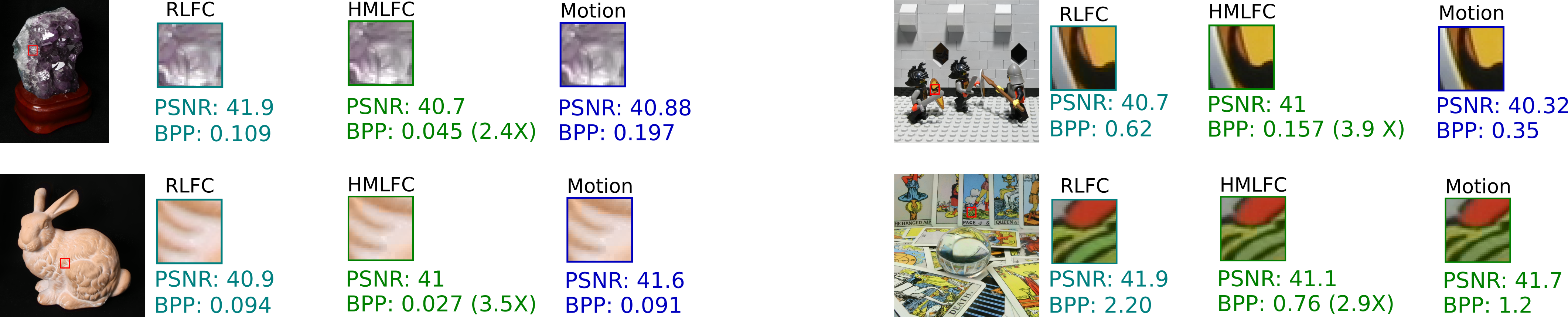}
 \caption{We show a zoomed in comparison between decoded images from the different schemes. A small region of size $32\times32$ marked in red box is selected and scaled upto $512\times512$ for visual quality comparison. The PSNR and bpp values for each of the methods are mentioned in the figure. For same PSNR values We find no additional visual degradation in HMLFC compared to RLFC and motion compensation. The factor of improvement of HMLFC over RLFC is highlighted in the bracket.}

 \label{fig:zoomed_in}
 \vspace*{-2.5em}
 \end{figure*}

\textbf{Future Work:} In the current implementation, we use per-pel motion compensation, i.e., a search for a matching block is performed at a pixel level. Using sub-pel motion compensation, i.e., sub-pixel level motion compensation to search for a matching block using bi-linear interpolation methods could provide better compression rates. We have implemented the hybrid approach for one specific hierarchical scheme (RLFC), and we would like to extend and test our hybrid approach for other hierarchical compression schemes that allow random access (e.g., Peter \& Stra{\ss}er~\cite{LFPeter01}). Adding depth information to the light field pixel data improves the rendering quality by a significant factor and provides more parallax. Extending our approach to compress depth information alongside image data is also a good direction for future work. Also, using motion compensation vectors for parallax correction to reduce artifacts during LF rendering may be possible.  Our current implementation focuses on 4D two-plane parameterization of the light fields; in the future, we would like to extend our compression approach to more complex parameterizations such as spherical~\cite{LFSphereihm1997}, panoramic~\cite{WelcomeLF}, and unstructured LF~\cite{LFdavis2012unstructured}. Our current implementation for encoding LFI is single threaded and slow. The encoding speed can be improved by a factor with a mutli-thread and parallelized implementation on the CPU or the GPU.